\documentstyle[12pt]{article}
\begin{document}
\begin{titlepage}

\title {Renormalization Group Equations for Seesaw Neutrino Masses.
 \thanks{~~Supported in part by the Polish Committee for Scientific
         Research under the grant 2 0165 91 01
and by the Swiss National Science Foundation.}}
\author{Piotr H. Chankowski
\thanks{On leave of absence from
Institute of Theoretical Physics, Warsaw University.}\\
Istituto Nazionale di Fisica Nucleare, Sezione di Padova\\
and\\
Dipartimento di Fisica "Galileo Galilei"\\
via F. Marzolo 8, 35131 Padova, Italy\\
\and
Zbigniew P\l uciennik$^{~\dagger}$\\
Institut f\"ur Theoretische Physik\\
Universit\"at Z\"urich\\
Sch\"onberggasse 9, 8001 Z\"urich, Switzerland\\
}

\date{June 1993}
\maketitle

\begin{abstract}
RGEs for coefficients of dim-5 operators giving rise to neutrino
masses in the seesaw mechanism
are written down in the SM, 2HDM and MSSM, and solved numerically.
RG evolution of these coefficients
modifies tree-level seesaw predictions for neutrino masses and
mixing angles in SO(10)-type GUT models as strongly as quark Yukawa
coupling evolution.
\end{abstract}

\begin{tabbing}
111111111111111111111111111111111111111111111111111111 \= 222222222 \kill
\> {\bf ZU-TH 20/93}\\
\> {\bf DFPD 93/TH/44}
\end{tabbing}

\end{titlepage}

\indent The MSW solution of the solar neutrino problem \cite{msw,lang}
has revived in the last few years interest in GUT model predictions for
the superlight neutrino masses generated via the seesaw mechanism
\cite{lang,inni}. Such models predict the neutrino mass matrix (sometimes up
to a normalization factor) using definite relations between quark Yukawa
couplings, Dirac type neutrino Yukawa couplings and Majorana mass matrix
for the right-handed neutrinos. In this letter we elaborate on
one technical aspect of such analyses: In comparing the GUT relations with
low energy charged fermion and neutrino masses one has to perform a RG
evolution of relevant parameters. The RG evolution of charged fermion
Yukawa couplings from the weak ($M_W$) to the unification scale ($M_X$)
is standard \cite{nasi} and has been included before
\cite{lang,inni}. The RG evolution of neutrino masses has so far been
neglected or treated inconsistently \cite{lang}. We furnish this
lacking element and demonstrate its possible numerical significance.

\indent  GUT models we are interested in contain  superheavy
 ~$SU(2)\times U(1)$ -singlet Majorana fields with a Dirac-type Yukawa
coupling to the SM neutrinos and Higgs boson.
To describe physics far below the GUT scale we
should decouple all heavy states and use an effective theory containing
light fields only.
Tree-graph exchange of superheavy right-handed neutrinos
gives rise, after integrating them out,
  to  dimension-5  operators coupling two
left-handed lepton doublets and two Higgses. Couplings of this
kind, being suppressed
by the inverse heavy mass factor, are neglected in the standard approach
in which one retains in the effective lagrangian
renormalizable interactions only. In order to obtain the seesaw neutrino
masses of the order ~${\cal O}(1/M_X)$ ~one must keep in the low energy
lagrangian also those nonrenormalizable higher dimension terms.
Similar couplings arise also in some string models \cite{lang} and
our RG analysis is also applicable in this case starting
from the scale where the effective theory is ~$SU_c(3)\times SU_L(2)
\times U_Y(1)$. ~We  write down one-loop RGEs for the coefficients of
the relevant dimension-5 operators which describe their evolution down
to ~$M_W$ ~scale. When Higgs fields are replaced by their VEVs this
coupling yields a superlight Majorana mass
term for the left-handed neutrinos.
In the Standard Model (SM) as well as the 2 Higgs Doublet Model (2HDM)
we need only one dimension-5 operator - ~$O_1$ . In the
SUSY case we have to include other operators related to
 ~$O_1$ ~by supersymmetry transformations because they mix under
renormalization. We solve numerically those RGEs in the SUSY case and
find that they introduce modifications to the tree-level seesaw formula
of the same order as the quark Yukawa coupling evolution.

\indent  While evolution of the neutrino masses themselves does not
provide very valuable information because the  overall scale of  heavy
Majorana mass is not fixed precisely in GUTs,  evolution of the neutrino
mixing angles (e.g. if a definite texture for the Majorana mass matrix is
assumed) may be important in comparing GUT predictions
with MSW solar neutrino problem solution.

\indent Superlight neutrino masses arise via the seesaw mechanism
after diagonalization of the mass term ($M\gg m$, ~generation indices
neglected):

\begin{eqnarray}
\Delta L_{mass} &=& -{1 \over 2}  \left( \matrix{ \nu & n } \right)
         \left( \matrix{ 0 & m
                     \cr m & M } \right)
         \left( \matrix{ \nu \cr n } \right)
\end {eqnarray}
 ( $\nu$ is the left-handed neutrino and $n$ is the left-handed
antineutrino).

\noindent It has two eigenvalues: ~$m_{heavy} \approx M$ ~and
 ~$m_{light}\approx {m^2 / M}$, ~with the light neutrino:
 ~$\nu_{light}=\cos\alpha~
\nu + \sin\alpha~ n$ ~where ~$\sin\alpha \approx m/M$.

\indent This viewpoint does not allow, however, any simple analysis
of the neutrino seesaw mass running. Moreover, a serious problem arise,
in theories (like MSSM) in which electroweak symmetry breaking proceeds
radiatively \cite{radbr}. On one hand, the tree-level VEV of the Higgs
boson(s) vanishes at $M_X$ scale and no neutrino mass is generated there.
On the other hand, to get Higgs VEVs one must use RGE of the effective
theory with all superheavy particles decoupled. As a result neutrinos stay,
at first sight, massless. The solution of the two above mentioned
problems emerges when one retains in the effective lagrangian higher
dimension operators generated by the decoupling procedure.
The heavy Majorana neutrino ~$n$ ~with the Yukawa coupling in the GUT
lagrangian of the form:
\begin{eqnarray}
\Delta L = - Y^{ab} n^a \epsilon_{ij} l_i^b H_j
\end {eqnarray}
(~$\epsilon_{ij}$ - antisymmetric ~SU(2)~ tensor;~a,b - generation
indices and
$H_i$ ~is the Higgs doublet transforming as (2, 1/2) under the
electroweak gauge group which in the 2HDM and MSSM gives masses to the
up-type quarks)
when integrated out, gives rise to the nonrenormalizable term
in the effective lagrangian
\begin{eqnarray}
\Delta L_{eff} = {1\over4}~ c_1^{ab} O_1^{ab}
\end{eqnarray}
where
\begin{eqnarray}
O_1^{ab} = (\epsilon_{ik} H_i l_k^a ) (\epsilon_{jl} H_j l_l^b)\nonumber
\end{eqnarray}
and ~$c_1^{ab}$ ~is a symmetric matrix of coefficients:
\begin{eqnarray}
c_1^{ab}(M_X) = 2~Y^{da} (M^{-1})^{dc} Y^{cb}\nonumber
\end{eqnarray}
(M is the Majorana mass matrix of $n$).
It is easy to see that when Higgs field is replaced by its VEV, (3) gives
rise to the neutrino mass term which up to ~${\cal O}(1/M^2_X)$~ is
equal to the mass term given by (1).

\indent RGE for coefficient ~$c_1$ ~gets contributions from the proper
vertex
corrections due to gauge boson exchanges and Higgs self coupling, and
from external wave function renormalization which introduces also the
dependence on Yukawa couplings.
In the 2HDM we get:

\begin{eqnarray}
{d\over dt} c^{ab}_1 &=&\left[{1\over2}\lambda_2 -3~g^2_2
+6 ~{\rm tr}\left(Y_u Y^{\dagger}_u\right)\right] c^{ab}_1\nonumber \\
&+&{1 \over2}{\left(Y_l Y^{\dagger}_l\right)}^{bc} c_1^{ca}
+{1 \over 2}{\left(Y_l Y^{\dagger}_l\right)}^{ac} c_1^{cb}
\end{eqnarray}
(~$t\equiv(4\pi)^{-2}\log(Q/M_W)$
 ~and summation over the generation index ~$c$ ~is understood ).

In the SM there is an additional contribution to the RHS (because the
Higgs boson couples to all fermions):
\begin{eqnarray}
\left[
6~{\rm tr}\left(Y_d Y^{\dagger}_d\right)
+2~{\rm tr}\left(Y_lY^{\dagger}_l\right)
\right]c^{ab}_1\nonumber
\end{eqnarray}
Where ~$Y_u, ~Y_d, ~Y_l$ ~are up-quark, down-quark and lepton
Yukawa matrices respectively and ~$\lambda_2$ ~is the Higgs self coupling:
\begin {eqnarray}
\Delta L_{higgs} &=& -{1\over 8} \lambda_2 \left(H^{\dagger} H \right)^2
\end{eqnarray}

\indent In the MSSM we
have to introduce other operators related to ~$O_1$ ~by SUSY because
they mix under renormalization due to diagrams in which gauge bosons
are replaced by the corresponding gauginos.
The relevant part of the effective
lagrangian reads:
\begin{eqnarray}
\Delta L_{eff} = {1\over 4}~c_1^{ab}~O_1^{ab}
           + c_{21}^{ab}~O_{21}^{ab}
           + c_{22}^{ab}~O_{22}^{ab}
           + {1\over 4}~c_3^{ab}~O_3^{ab}
\end{eqnarray}
with
\begin{eqnarray}
O_{21}^{ab} = (\epsilon_{ik} h_i L_k^a ) (\epsilon_{jl} H_j l_l^b)\nonumber
\end{eqnarray}
\begin{eqnarray}
O_{22}^{ab} = (\epsilon_{ik} h_i l_k^a ) (\epsilon_{jl} H_j L_l^b)\nonumber
\end{eqnarray}
\begin{eqnarray}
O_3^{ab} = (\epsilon_{ik} h_i L_k^a ) (\epsilon_{jl} h_j L_l^b)\nonumber
\end{eqnarray}

\noindent ~$h$ ~is a higgsino  and ~$L^a$ ~is a slepton.
Operators ~$O_1$ ~to ~$O_3$ ~form a basis
in the space of operators which mix under renormalization.

\indent For a superpotential containing the singlet neutrino superfield
 ~$\hat N^a$:
\begin{eqnarray}
w &=&{1\over 2}~M^{ab} \hat N^a \hat N^b
     +Y^{ab} \hat N^a \epsilon_{ij} \hat L_i^b \hat H_j
\end{eqnarray}
we get after integrating out ~$n^a$ ~and its superpartner:

\begin{equation}
{1 \over 2} c_1^{ab} =
  c_{21}^{ab} =
  c_{22}^{ab} =
{1 \over 2} c_3^{ab} =
Y^{ca} (M^{-1})^{cd} Y^{db}
\end{equation}

\noindent The RGE for ~$c_1^{ab}$ ~in this case reads:
\begin{eqnarray}
  {d \over dt} c^{ab}_1 &=&
  \left[{1\over2} \lambda_2-3 g^2_2 + G^2_{1H} +3 G^2_{2H}\right.\nonumber\\
  &+&\left.{1\over2}G^2_{1L} + {3\over2}G^2_{2L}
  + 6~tr\left(Y_uY^{\dagger}_u\right)\right] c^{ab}_1 \nonumber\\
  &+& {1 \over 2}\left[\left(Y_{lHe} Y^{\dagger}_{lHe} \right)^{bc}
  +\left(Y_{lhE} Y^{\dagger}_{lhE} \right)^{bc} \right] c^{ca}_1
 \\
  &+& {1 \over 2}\left[ \left(Y_{lHe} Y^{\dagger}_{lHe} \right)^{ac}
+ \left(Y_{lhE} Y^{\dagger}_{lhE} \right)^{ac}\right] c^{cb}_1\nonumber \\
  &-&\left(2~G_{1H}G_{1L}+6~G_{2H}G_{2L}\right)
  \left(c^{ab}_{21} + c^{ba}_{21}\right)\nonumber \\
  &-&\left(2~G_{1H}G_{1L}+2~G_{2H}G_{2L}\right)
  \left(c^{ab}_{22} + c^{ba}_{22}\right)\nonumber
\end{eqnarray}
\\
In equation (9) ~$G_{1L},~G_{2L},~G_{1H},~G_{2H}$
 ~are ~U(1) ~and ~SU(2) ~gaugino couplings to leptons and Higgs boson.
$Y_{lHe}$ and $Y_{lhE}$ are Yukawa couplings of leptons to Higgses and
higgsinos respectively.
In the strict SUSY limit they are all equal to the corresponding gauge
and Yukawa couplings ~$g_1$, ~$g_2$ and $Y_l$.
 ~In the same limit ~$\lambda_2
=g^2_1+g^2_2$. ~We display, however, this equation in its more
general form which allows one to decouple (when necessary)
all heavy sfermions in one collective threshold. The RGEs for other
MSSM couplings written in the same general form can be found in \cite{my}.\\
 \noindent The remaining RGEs read:

\begin{eqnarray}
  {d \over dt} c^{ab}_3 &=&\left[2 g_1^2+2 g^2_2
  +6~tr\left(Y_u Y^{\dagger}_u\right)\right]c^{ab}_3\nonumber\\
  &+&\left(Y_lY^{\dagger}_l\right)^{bc} c^{ca}_3
  +\left(Y_l Y^{\dagger}_l\right)^{ac} c^{cb}_3\\
  &-&\left(2~g_1^2 +6~g_2^2 \right)
  \left(c^{ab}_{21}+c^{ba}_{21}\right)\nonumber \\
  &-&\left(2~g_1^2 +2~g_2^2 \right)
  \left(c^{ab}_{22}+c^{ba}_{22}\right)\nonumber
\end{eqnarray}

\begin{eqnarray}
  {d \over dt} c^{ab}_{21} &=&\left[4~g^2_2 - 2~g_1^2
  +6~tr\left(Y_u Y^{\dagger}_u\right)\right]c^{ab}_{21}
  + 2~g^2_2 ~c^{ba}_{21}\nonumber\\
  &+&\left(Y_l Y^{\dagger}_l\right)^{bc} c^{ca}_{21}
  +\left(Y_l Y^{\dagger}_l\right)^{ac} c^{cb}_{21}\nonumber \\
  &+&\left(g_1^2-g_2^2\right)\left(c_{22}^{ab}+c_{22}^{ba}\right)\\
  &-&{1\over 2}\left(g_1^2+5~g_2^2 \right)
  \left(c_1^{ab}+ c_3^{ab}\right)\nonumber
\end{eqnarray}
 \newpage

\begin{eqnarray}
  {d \over dt} c^{ab}_{22}&=&\left[-4~g_1^2-2~g^2_1
  +6~tr\left(Y_u Y^{\dagger}_u\right)\right]c^{ab}_{22}
  +2~g^2_2~c^{ba}_{22}\nonumber\\
  &+&\left(Y_l Y^{\dagger}_l\right)^{bc} c^{ca}_{22}
  +\left(Y_l Y^{\dagger}_l\right)^{ac}c^{cb}_{22}\nonumber \\
  &+&\left(g_1^2-g_2^2\right)\left(c_{21}^{ab}+c_{21}^{ba}\right)
  -4~g_2^2~c^{ab}_{21}\\
  &-&{1\over2}\left(g_1^2-g_2^2 \right)
  \left(c_1^{ab}+ c_3^{ab}\right)\nonumber
\end{eqnarray}

In (10-12) we implement all SUSY relations between couplings
but we retain the distinction between ~$c^{ab}_{2i}$ ~and ~$c^{ba}_{2i}$
 ~because in principle one could have a model where they are not symmetric.
In (4) and (9-12) there is no mixing with operators involving the other
Higgs doublet ~$(2,-1/2)$~ because there is no appropriate coupling in the
MSSM and simple 2HDMs.
Contributions from Yukawa couplings come only from
wave function renormalization and therefore do not cause
operator mixing. This fact simplifies considerably the analysis
of lepton mixing angle evolution presented below.

\indent In order to demonstrate the potential importance of the RGE
evolution for neutrino masses and mixing anles
we have solved  equations (9-12) for the SUSY case numerically,
evolving simultaneously gauge couplings and Yukawa couplings in
one-loop approximation and
including nonlinearities from the ~${\rm3}^{\rm rd}$ ~family. We take
 ~$\alpha_s (M_Z) = 0.12$ ~and ~$\sin^2\theta_W = 0.233$ ~and allow
 ~$Y_{top}(M_X)$ ~to vary from  0.2 to 6. Our unification scale is
 ~$M_X=10^{16}$ GeV.  We consider two distinctly different cases:\\
\noindent $i)$ SO(10)-type unification for Yukawa couplings with

$Y_{t} (M_X) = Y_{b}(M_X)= Y_{\tau}(M_X)$ \\
\noindent $ii)$ An example of ~SU(5)-type Yukawa coupling unification with

$Y_{t} (M_X) = 10 ~Y_{b}(M_X)=10 ~Y_{\tau}(M_X)$ . \\
\noindent After evolving the ~$Y$s down to ~$M_Z$ ~we fix ~$\tan\beta$
 ~fitting $m_\tau$=1.784 GeV:
 ~$m_\tau(M_Z)=\sqrt{2}\left(Y_{\tau}(M_Z)/g_2\right) M_W \cos\beta$
 ~and calculate
 ~$m_{top}(M_Z)$ ~and ~$m_{b}(m_b)$ ~which is presented in Fig.1a .
$Y_t(M_X)$~as a function of $m_{top}$ is presented in Fig.1b.
This sets
the background for the evolution of coefficients ~$c_i$ and allows one to
assess the proximity of the Landau pole.
 ~Our aim in this note is to demonstrate the effects of the neutrino
masses and mixing angles evolution rather than performing a detailed
study of Yukawa coupling unification. Therefore, we neglect possible
deviations from the assumed relations between Yukawa couplings at ~$M_X$
 ~(threshold corrections) which could result in  ~$m_b$ ~closer to
its experimental value  ~$m_b(m_b)=4.25\pm 0.10$~ GeV \cite{kwarki}
\footnote{We neglect also possible effects of decoupling  superpartners
at some scale ~$M_{SUSY} = {\cal O}(1~TeV)$ ~on the RG evolution.}.
This simplified approach suggests
(see \cite{lanpol} for a detailed analysis) that Yukawa
coupling unification favours a heavy top quark.

\indent We work in the approximation of small ~${\rm 3}^{\rm rd}$
 ~generation mixing.
In evolving ~$Y_{t} , ~Y_{b} , ~Y_{\tau} $ ~(or eg. ~$Y_u , ~Y_d ,
 ~Y_e$ ) we
 neglect this mixing altogether and in evolving quark and neutrino
mixing angles we use approximate RGEs \cite{nasi,berger} which for the MSSM
read:
\begin{eqnarray}
 {d \over dt} \theta_{13}^{\nu} &=&
  - ~Y_{\tau}^2~\theta_{13}^{\nu}
  \nonumber\\
 &~& \\
 {d \over dt} \theta_{13}^{KM} &=&
  - ~(Y_{t}^2+Y_b^2)~\theta_{13}^{KM}
\nonumber
\end{eqnarray}
Those simple RGEs arise because Yukawa couplings responsible for
lepton mixing angle evolution appear exactly in the same way
in (9-12). The mixings between different operators ~$O_1$ -- $O_3$~
and different generations can be factorized substituting
 ~$c^{ab}_i\equiv A^{ab}~b_i$~ and, in fact, one could
guess (13) without calculating (9-12) explicitly.
The results are presented in Fig.2a -- 2d.
In all figures there are 2 sets of curves: for Yukawa unification
$i)$ - solid lines and $ii)$ - dashed ones.\\
\noindent Fig.2a presents running of the coefficient ~$c_1$
 ~for the ~${\rm 3}^{\rm rd}$ and for
the ~${\rm 1}^{\rm st}$ (or ~${\rm 2}^{\rm nd}$)
generation.
 For heavier top, close to the fixed point of the RGEs, Yukawa
contributions  overcome the gauge coupling contributions and change
the direction of evolution of coefficients ~$c_i$, ~just like
in the case of quark Yukawa couplings.\\
\noindent
Fig.2b presents running of the neutrino  1-3 (or 2-3) mixing angle, which is
sizeable only for large ~$Y_{\tau}$ - i.e. in the case of
 ~SO(10)-type of Yukawa couplings unification  and large $m_{top}$.\\
\noindent Fig.2c  presents the joint effect
of both KM and neutrino mixing angle evolution. If we
assume that they are equal at ~$M_X$ ~as predicted by some models
\cite {lang} then their ratio at ~$M_W$, ~which is plotted in Fig.2c,
is due only to RG running .
Running of the neutrino mixing angle  is
of the same order of magnitude as KM angle running  and partly cancels
effects of the latter in
 ~$\theta_{13}^{\nu}/ \theta_{13}^{KM}$.
 One can see that evolution even slightly
strengthens the discrepancy
 between ~$\theta^{KM}_{13}=(0.3 -1.0)\times10^{-2}$
 ~and the small angle MSW solar neutrino problem solution:
 ~$\theta^{\nu}_{13}=(3-5)\times 10^{-2}$ ~\cite{lang}. (In SUSY-GUT seesaw the
solar neutrino problem is solved by ~$\nu_e \rightarrow \nu_{\tau}$
 ~transition.)\\
\noindent Finally in Fig.2d we present the values of coefficient ~$\eta$~
incorporating the effects of both neutrino and quark mass running
in the seesaw formula:
\begin{eqnarray}
m_{\nu}^a (M_Z) &=& {\eta}^a {{m^{a~2}_{up} (M_Z)} \over M(M_X)}
\end{eqnarray}
(no sum over generation index  a) \\
 ~$\eta_i$ ~for ~$i=1,2$ ~are independent of ~$m_{top}$
 ~because in these quantities trace--type Yukawa coupling contributions
to the evolution of $Y_{top}$ and $c_1$ cancel each other. For
 ~$\eta_3$ ~the trace-type contributions (~$\sim tr(Y_t Y_t^{\dagger})$~)
to the evolution of
 ~$c_1$ ~and ~$Y_t$ ~cancel each other as well but the evolution of ~$Y_t$ ~due
to the non-trace
contributions (~$\sim (Y_t Y_t^{\dagger})$~) is stronger than analogous
evolution of ~$c_1$ ~(~$\sim (Y_{\tau} Y_{\tau}^{\dagger})$~) and results
in a large ~$\eta_3$ ~for a heavy top quark.
Values of coefficients ~$\eta$ ~have been given in \cite{lang}
(with light quark masses renormalized at lower scales). Even though
numerical differences ($\sim 20\%$) are insignificant considering our
ignorance of the overall scale of M we believe our results are more complete.
We include carefully the effective dim-5 coupling running and
state the assumptions concerning the charged fermion Yukawa coupling
evolution more clearly, including
assumed Yukawa coupling unification - the essential
element of seesaw neutrino mass predictions.

\indent To summarize,
 superlight neutrino masses arising via the seesaw mechanism can
be most naturally viewed as a manifestation of an effective dimension-5
operator. This approach is the only one available in models with
radiative ~SU(2)$\times$U(1) ~symmetry breaking. The RG analysis based
on dimension-5 operators provides a systematic and unambiguous method of
incorporating radiative corrections to the GUT predictions for neutrino
seesaw masses.
Running of the coefficients ~$c_i$ ~is an as important element of neutrino
mass predictions from GUTs as quark Yukawa coupling running.
For a heavy top quark the most important contribution to RGEs comes
from Yukawa couplings of the  ~${\rm 3}^{\rm rd}$ generation.
 ~Any specific model predicting neutrino mixing angles from GUT seesaw
must take into account equations (9-12) when it is compared with experiment.\\

\indent We thank Prof. S.Pokorski and Dr.  M.Olechowski for discussions which
 initiated
this study. Z.P. would like to thank Prof. D.Wyler for kind hospitality
extended to him during his stay in Z\"urich.

\newpage

{\bf FIGURE CAPTIONS}
\vskip 0.5cm

\noindent {\bf Figure 1.}~
{\bf a)}~$m_b(m_b)$ ~and ~{\bf b)}~$Y_t(M_X)$ ~shown as a function of
{}~$m_{top}$
 ~in unification schemes ~$i)$ ~and ~$ii)$.

\vskip 0.5cm
\noindent {\bf Figure 2.}~
The result of the dim-5 coupling evolution: ~{\bf a)}~ coefficients ~$c_1$
 ~for the
 ~${\rm 1}^{\rm st}$ ~and ~${\rm 3}^{\rm rd}$ ~generation and ~{\bf b)}~ lepton
 mixing
angle ~$\theta^{\nu}_{13}$. ~Joint effect of dim-5 coupling and quark Yukawa
coupling evolution: ~{\bf c)} ~$\theta_{13}^{\nu}(M_W)/ \theta_{13}^{KM}(M_W)$
 ~and ~{\bf d)}~ coefficient
 ~$\eta$ ~for the  ~${\rm 1}^{\rm st}$ ~and ~${\rm 3}^{\rm rd}$ ~generation.

\end{document}